\documentclass[11pt]{article}
\usepackage{epsfig}
\usepackage{xspace}

\def\cs{{\it Cloud Scheduler}\xspace}
\def\js{{\it Condor Job Scheduler}\xspace}

\begin{document}

\begin{center}
{\large\bf
Cloud Scheduler: 
a resource manager for distributed compute clouds} 
\end{center}
\vskip 20pt

\begin{center}
P. Armstrong$^1$,
A. Agarwal$^1$,
A. Bishop$^1$,
A. Charbonneau$^2$,
R. Desmarais$^1$,
K. Fransham$^1$,
N. Hill$^3$,
I. Gable$^1$,
S. Gaudet$^3$,
S. Goliath$^3$,
R. Impey$^2$,
C. Leavett-Brown$^1$,
J. Ouellete$^3$,
M. Paterson$^1$,
C. Pritchet$^1$,
D. Penfold-Brown$^1$,
W. Podaima$^2$,
D. Schade$^3$,
R.J. Sobie$^{1,4}$
\end{center}

\begin{center}
$^1$ Department of Physics and Astronomy, 
University of Victoria, Victoria, Canada V8W 3P6\\
$^2$ National Research Council Canada,
100 Sussex Drive, Ottawa, Canada \\
$^3$ 
National Research Council Herzberg Institute of Astrophysics, 
5071 West Saanich Road, Victoria, BC, Canada, V9E 3E7 \\
$^4$ Institute of Particle Physics of Canada.
\end{center}

\begin{abstract}
The availability of Infrastructure-as-a-Service (IaaS) computing clouds gives
researchers access to a large set of new resources for running complex 
scientific applications.
However, exploiting cloud resources for large numbers of jobs requires 
significant effort and expertise.
In order to make it simple and transparent for researchers to deploy 
their applications, we have developed a virtual machine resource manager 
(\cs) for distributed compute clouds.
\cs boots and manages the user-customized virtual machines in response
to a user's job submission.
We describe the motivation and design of the \cs and 
present results on its use on both science and commercial clouds.
\end{abstract}

\section{Introduction}

Infrastructure as a Service (IaaS) cloud computing is emerging as a new 
and efficient way to provide computing to the research community.
Clouds are considered to be a solution to some of the problems
encountered with early adaptations of grid computing where
the site retains control over the resources and the user must 
adapt their application to the local operating system, software and policies.
This often leads to difficulties especially when a single resource provider 
must meet the demands of multiple projects or when projects cannot conform
to the configuration of the resource provider.
IaaS clouds offer a solution to these challenges by delivering computing 
resources using virtualization technologies.
Users lease the  resources from the provider and install their 
application software within a virtual environment.
This frees the providers from having to adapt their systems to 
specific application requirements and removes the software constraints 
on the user applications.
In most cases, it is easy for a user or a small project to create their 
virtual machine (VM) images and run them on IaaS clouds.   
However, the complexity rapidly increases for projects with large user 
communities and significant computing requirements.
In this paper we describe a system that simplifies the use of IaaS clouds
for High Throughput Computing (HTC) workloads.

The growing interest in clouds can be attributed in part to the ease in 
encapsulating complex research applications in Virtual Machines (VMs), often
with little or no performance degradation \cite{hepix-vm-benchmark}.
Studies have shown, for example, that particle physics application code 
run equally well in a VM or on the native system \cite{chep-vm}.
Today, open source virtualization software such as Xen \cite{xen} and 
KVM \cite{kvm}
are incorporated into many Linux operating system distributions,
resulting in the use of VMs for a wide variety of applications.
Often, special purpose servers, particularly those requiring high
availability or redundancy, are built inside a VM making them independent
of the underlying hardware and allowing them to be easily moved or replicated.
It is also not uncommon to find an old operating system lacking the
drivers for new hardware, a problem which may be resolved by running
the old software in a virtual environment.

The deployment and management of many VMs in an IaaS cloud 
is labour intensive.
This can be simplified by attaching the VMs to a job scheduler
and utilizing the VMs in a batch environment.
The Nimbus Project \cite{nimbus} has developed the {\it one-click cluster} 
solution.
This provides a batch system on multiple clouds using one type of VM
\cite{one-click-cluster}.

We further simplify the management of VMs in an IaaS cloud and provide
new functionality with the introduction of \cs.
\cs provides a means of managing user-customized VMs on any number 
of science and commercial clouds\footnote{Commercial cloud providers 
include Amazon, RackSpace and IBM, to name a few.
Science clouds use hardware resources funded by governments for research 
purposes and are located in universities or national laboratories.}.

In the following sections we present the architecture of our system 
and highlight the role of \cs to manage VMs on IaaS clouds in a 
batch HTC environment.
We present early results on the operation, and highlight the successes
and issues of the system.
We summarize the paper with a discussion of the future developments.

\section{System architecture}

\begin{figure*}
\centering
\epsfig{file=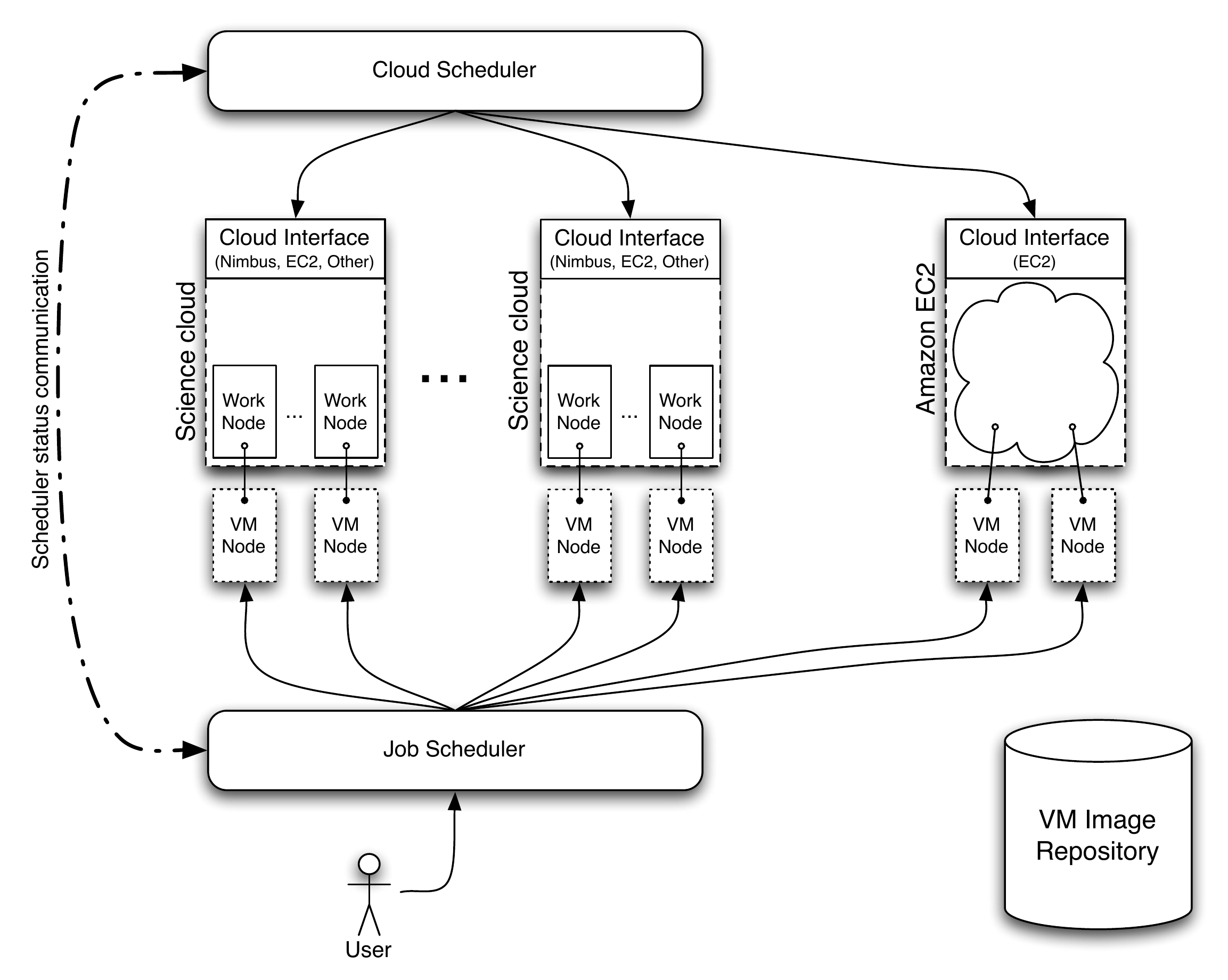, width=15cm }
\caption{An overview of the architecture used for the system.
A user prepares their VM image and a job script.  
The job script is submitted to the \js.
The \cs reads the job queue and makes a request to boot the 
user VM on one of the available clouds.
Once there are no more user jobs requiring that VM type, the \cs makes 
a request to the respective cloud to shutdown the user VM.
}
\label{fig:overall_arch}
\end{figure*}

The architecture of the HTC environment running user-customized 
VMs on distributed IaaS clouds is shown in fig.~\ref{fig:overall_arch}.
A user creates their VM and stores it in the {\it VM image repository}.
They write a job script that includes information about their VM
and submits it to the \js.     
The \cs reads the queues of the \js, requests that one of the available
cloud resources boot the user VM,  the VM advertises itself to the 
\js which then dispatches the user job to that VM.

In this section we describe the following components: 
{\it VM image repository}, the cloud resources and the \js.
The \cs is discussed in more detail in the following section.

\begin{description}

\item {\bf VM image repository} \\
The user  builds their VM by first retrieving a base image from
a {\it VM image repository}.
The base images are simple Linux VM images or images that
include project or application based code.
Once the user has modified their image, they will store it back in
the {\it VM image repository}.
The repository may reside at a single site or be distributed,
however, it must be accessible to the cloud resources.

\item {\bf Cloud resources} \\
The system currently supports Amazon EC2 and IaaS clouds using Nimbus \cite{nimbus}.
Support for IaaS clouds using OpenNebula \cite{open-nebula} and 
Eucalyptus \cite{eucalyptus} is under development.

\item {\bf Job Scheduler} \\
The job scheduler used in the system is the Condor HTC job scheduler \cite{condor}.
Condor was chosen because it was designed to utilize 
heterogeneous idle workstations which makes it ideal to use as a 
job scheduler for a dynamic VM environment.
Condor has a central manager which matches user jobs to resources 
based on job and resource attributes.
The Condor {\it startd daemon} must be installed and started when a 
VM image is booted. 
The VM then advertises its existence to the Condor central manager\footnote{We 
use Condor Connection Brokering (CCB) to allow VM worker nodes that use
Network Address Translation (NAT) to connect to the Condor Central 
Manager.}.
Users submit jobs by issuing the \texttt{condor\_submit} command.
The user must add a number of additional parameters specifying the 
location and properties of the VM.   
The description of the parameters is found in Appendix I.

\end{description}

\section{Cloud Scheduler}

The \cs is an object oriented python-based package designed to 
manage VMs for jobs based on the available cloud resources and job requirements.     
Users submit jobs to the \js after they have been 
authenticated using X.509 Proxy Certificates \cite{rfc3820}.
The certificates are also used to authenticate 
starting, shutting down, or polling VMs with Nimbus clusters. 
Authentication with EC2 is done by using a standard shared access key and 
secret key.

In the following subsections, we describe the \cs's object classes
and how they are used to manage the VMs for the jobs.
Finally we discuss the job scheduling and load balancing considerations.


\subsection{Resource and job management classes}

\cs keeps track of jobs and resources with a set of cloud and 
job management classes (see Tables~\ref{table:cloudclasses} and  
\ref{table:jobclasses}).
The cloud management classes include the {\it ResourcePool}, 
{\it Cluster} and {\it VM} classes.
The {\it ResourcePool} is a list of cloud resources that is read on 
initialization, but can be updated at run-time. 
The {\it Cluster} class contains static information describing 
the properties of each cloud and a dynamic list of {\it VM} objects 
running on that cloud. 
The {\it VM} class contains  information describing the properties
and state of a VM. 

The job management classes include the {\it JobPool} and {\it Job} classes.
The {\it JobPool} class contains a list of job objects that 
are derived from the jobs submitted to Condor.
The {\it Job} class contains the properties of the user job.


\subsection{VM management}

When \cs is started it reads the general and cloud configuration files. 
It starts the following threads that are run on a periodic basis.

\begin{enumerate}
\item 
The {\it JobPoller} thread maintains the state and metadata 
of the jobs that are queued and running on the \js. 
It effectively maps the \js queue into the {\it JobPool}.

\item
The {\it Scheduler} thread starts and stops VMs based on the information 
in the {\it JobPool}, satisfying the resource demands of the workload. 
The design goal of \cs is to leave prioritization and scheduling decisions 
in the domain of the \js. 
However, the order in which the {\it Scheduler} thread provides resources can 
impact the scheduling algorithms of the \js. 
Job scheduling and load balancing are discussed in more detail in the following section. 

The {\it Scheduler} thread also monitors the VMs and, if necessary, updates the 
state of VMs in the {\it JobPool}.  
It will shut down VMs that are in an error state;  if there are jobs
that still require this VM, then the {\it Scheduler} will  start a new
instance of the VM to replace the one it has shut down.

\item
The {\it CleanUp} thread stops VMs that are no longer required.
It can correct the state of the job in the {\it JobPool}.
If a VM is shut down due to an error, then the {\it CleanUp} thread
changes the state of the job in the {\it JobPool} from ``scheduled''
to ``new'' so that a new VM can be created for that job.

\end{enumerate}

When \cs is shut down, it can either shut down all the VMs that is has started,
or it can persist its state. 
In the latter case, the VMs continue to run the jobs.
\cs reloads the state when it is restarted and resumes managing the jobs
and resources.


\subsection{Job scheduling and load balancing}

The \js is designed to manage job prioritization and scheduling 
\cite{condor}.
As mentioned, \cs can impact Condor's job scheduling. 
For example, consider two queued jobs, the first submitted job requires a 
VM type of VM-A, and a second submitted job requires a VM type of VM-B. 
If Cloud Scheduler starts a VM-B first (as the resources to boot a VM-B 
are available), then jobs requiring VM-B will run before jobs requiring VM-A.

\cs can be configured to take into account user fairness, resource utilization, 
and user priorities. 
Currently, \cs will start as many VMs that will fill the resources. 
The VMs are evenly distributed among all users with jobs in the queue. 
As other users submit jobs, \cs re-balances the VM 
allocations by shutting down over-allocated VMs and starting under-allocated VMs. 
For example, a single user will get the full allocation of \cs's VMs, 
but once a second user submits jobs, half of the first user's VMs 
will be shut down to free resources for the second user. 

\cs can re-balance the VM distribution by shutting down 
VMs gracefully or by killing them outright.  
When configured for graceful shutdown, \cs switches the Condor-state of pending jobs 
requiring the over-allocated VMs to a ``held'' state  
thus preventing them from being dispatched to currently running VMs. 
The next VM of the over-allocated type to finish its job can be 
safely shutdown without affecting running jobs. 
When \cs is configured to kill VMs outright, the VMs are shutdown immediately 
without waiting for the job to finish. 
If a VM is killed while a job is running, then the \js  re-queues 
the job for execution.

\begin{table}
\begin{center}
\caption{\label{table:cloudclasses}  
The cloud management classes}
\vspace{0.25cm}
\begin{tabular}{|ll|} \hline
\multicolumn{2}{|l|}{\bf ResourcePool class} \\
Object      & Description \\ 
ClusterList & The list of {\it Cluster} objects \\
\hline \hline
\multicolumn{2}{|l|}{\bf Cluster class} \\
Object      & Description \\ \hline 
name        & The name of cluster \\
host        & The hostname of cluster  \\
cloud\_type & The type of IaaS software (Nimbus, Eucalyptus, etc) \\
memory      & The RAM available for a VM \\
cpu\_archs  & The CPU architectures available  \\
networks    & The network types available (private or public or both) \\
vm\_slots   & The maximum number of VMs allowed on the cluster \\
cpu\_cores  & The maximum number of cpus allowed for a single VM \\
storage     & The scratch space available \\
vms         & The {\it VM List} \\ 
\hline \hline
\multicolumn{2}{|l|}{\bf VM class} \\
Object      & Description \\ \hline 
name        & The name assigned to the VM \\
id          & The cluster-specific identifier  \\
vmtype      & The type of the VM  \\
vmstate     & The state of the VM (Starting, Running or Error) \\
hostname    & The hostname of the VM   \\
clusteraddr & The address of the IaaS head node that controls this VM \\
network     & The type of networking of the VM  (private or public) \\
cpuarch     & The cpu architecture of the VM (i386 or x86\_64) \\
image       & The VM image used to boot the VM \\
memory      & The VM RAM \\
cpucores    & The number of cores in the VM  \\
storage     & The size of the scratch space used by the VM \\
errorcount  & The number of times the VM has given an error response \\
lastpoll    & The date/time of the latest update \\
last\_state\_change & The date/time of the last VM state change \\ 
\hline 
\end{tabular}
\end{center}
\end{table}

\begin{table}
\begin{center}
\caption{\label{table:jobclasses}  
The job management classes}
\vspace{0.25cm}
\begin{tabular}{|ll|} \hline
\multicolumn{2}{|l|}{\bf JobPool class} \\
Object        & Description \\ \hline 
NewList       & List of new Job objects \\
ScheduledList & List of scheduled Job objects \\ 
\hline \hline
\multicolumn{2}{|l|}{\bf Job class} \\
Object      & Description \\ \hline 
GlobalJobID & The Condor job ID \\
User        & The user that submitted the job  \\
Priority    & The priority given in the job submission file (default = 1) \\
VMType      & The VMType  \\
VMNetwork   & The network required  (private/public) \\
VMCPUArch   & The CPU architecture(x86 or x86\_64) \\
VMName      & The name of the image the job is to run on \\
VMLoc       & The location (URL) of the image \\
VMAMI       & The Amazon AMI of the image \\
VMMem       & The amount of memory in MB  \\
VMCPUCores  & The number of cpu cores  \\
VMStorage   & The amount of storage space \\
\hline 
\end{tabular}
\end{center}
\end{table}

\section{Results}

The system is currently being used for particle physics and astronomy applications.
At the moment we operate two independent systems for each research community.
In addition, we are commissioning a cluster at the University of Victoria that will 
be shared by a number of groups.

\subsection{Particle physics}
The system is being used to generate simulated particle physics events for
the BaBar experiment based at the Stanford Linear Accelerator Center (SLAC).
A number of faculty at the University of Victoria are members of the
collaboration and one of the responsibilities of the group is to use Canadian
computing resources for the generation of the simulated data.
Up to now we have used standalone facilities and also a grid of
facilities \cite{gridx1, babargrid}.

The simulation application contains C++ and FORTRAN code.
The current size of the VM is approximately 16 GBytes and requires about
1 GB of RAM.
The simulation requires access to calibration databases that vary in size
up to 2 TB.
The application accesses the databases on a modest but regular basis.
The databases can be remote if the network connectivity is good.
In Canada, we are able to use the CANARIE network which connects research
and educational institutions with a multi-gigabit network \cite{canarie}.
Our link to Amazon EC2 is through the commodity network and for the time being,
we have a copy of the databases on Amazon Storage to overcome the network
limitation.
Typically the simulation requires 6 hours and  produces an
output file of 100GB.
All output files are copied back to the University of Victoria where they
are merged and sent to SLAC.

The simulation production is operated by a single expert user.
There are no other users of the cloud system.
The user prepares the job scripts and submits them to the \js.
The \cs manages 80 VMs provided by three clouds located
at the University of Victoria, the National Research Council (NRC) 
of Canada in Ottawa and the Amazon EC2 cloud in the eastern US.
We limit the number of EC2 VMs due to the cost, however, we are
exploring the use of the EC2 on a variable basis dependent on the price.

The system is performing very well completing more than 2000 seven-hour
jobs in approximately one week.
When jobs are submitted to the queue, the \cs successfully boots as 
many VMs as are required. 
Once booted, the VMs successfully identify themselves to the job scheduler 
and begin to run jobs in the queue.  
As expected, the three cloud sites become a single distributed cloud, 
and simulation production proceeds identically as on a traditional cluster.

Issues that arose during the run were mainly centered around database access.  
To run efficiently, the jobs require fairly fast, low latency access to the databases.  
When the databases are hosted at the UVic site or the NRC site,
the database access is over the CANARIE network described above, 
and the jobs are limited by the speed of the CPU rather than the I/O.  
However, when the jobs run on Amazon EC2, the network bandwidth between EC2 and the
other two cloud sites is not sufficient for the jobs to run efficiently, and 
the jobs take approximately double the time to run to completion.  
This was solved by hosting a copy of the databases on Amazon Simple Storage 
System (S3) and accessing that copy from jobs running on EC2.

Another issue involved the quality of the data being produced on EC2.  
The BaBar collaboration places strict checks on simulation production; 
any remote site that produces data for the collaboration must pass a series 
of tests that compare data produced at the remote site to a set of reference 
data produced at SLAC.  
When running on standard EC2 instances with older AMD CPUs, the data that 
was produced was different enough from the reference data
to be unacceptable to the collaboration.  
Running the jobs on Amazon's ``high CPU'' instances
with newer Intel chips solved this problem.

\subsection{Astronomy}
The system is also a central component of CANFAR \cite{canfar}, an astronomy project 
led by researchers at the University of Victoria and the National Research Council of 
Canada Herzberg Institute of Astrophysics. 
The goal is to provide researchers the ability to create custom environments for analyzing data
from survey projects. 
Currently the system is available to early adopters, using two clusters: one 
at the WestGrid facility located at the University of Victoria (25 machines, 200 cores) 
and one at the Herzberg Institute of Astrophysics (6 machines, 32 cores). 
The system has been tested successfully with over 9,000 jobs utilizing more 
than 33,000 core hours. 
Work is currently underway to streamline the process of creating and sharing 
virtual machines with researchers, to allow them to easily make VM images that suit 
their needs and ready for deployment on cloud resources.

\section{Summary}

We have presented a new method for running large scale 
complex research applications on IaaS computing clouds. 
The development of \cs simplifies the management of 
virtual machine resources in a distributed cloud environment 
\cite{cloudscheduler} by hiding the complexity of VM management.
We have demonstrated that the system works for both
astronomy and particle physics applications 
using multiple cloud resources.
We have shown that the system is robust and fault tolerant.
We described our plans to further develop \cs and address other
issues for running HTC workloads in a cloud environment.

The support of CANARIE, the Natural Sciences and Engineering Research Council,
the National Research Council of Canada and Amazon are acknowledged.

\newpage
\noindent
{\Large\bf Appendix I: Job submission customization}
\vspace{0.5cm}

The user job submission script follows the Condor format, however,
a number of custom attributes are required to configure the VM.
The following table lists the additional items.

\vspace{0.5cm}
\begin{center}
\begin{tabular}{|ll|}  \hline
Attribute       & Description \\ \hline \hline
VMType          & Unique name of required VM \\ 
VMLoc           & URL (Nimbus) of the image \\
VMAMI           & AMI (EC2-like) of the image \\
VMCPUArch (x86) & CPU architecture (x86 or x86\_64) \\
VMCPUCores (1)  & Number of CPU cores  \\
VMStorage       & Required storage space  \\
VMMem           & RAM required \\
VMNetwork       & Network required  (public/private) \\ \hline
\end{tabular}
\end{center}

The {\it VMType} is a custom attribute of the VM advertised to the Condor 
central manager. 
It is specified on the VM condor\_config or condor\_config.local file. 

A sample job script is listed below.

\begin{verbatim}
Regular Condor Attributes
Universe                = vanilla
Executable              = script.sh
Arguments               = one two three
Log                     = script.log
Output                  = script.out
Error                   = script.error
should_transfer_files   = YES
when_to_transfer_output = ON_EXIT
#
# Cloud Scheduler Attributes
Requirements = 
+VMType                 = "vm-name"
+VMLoc                  = "http://repository.tld/your.vm.img.gz"
+VMAMI                  = "ami-dfasfds"
+VMCPUArch              = "x86"
+VMCPUCores             = "1"
+VMNetwork              = "private"
+VMMem                  = "512"
+VMStorage              = "20"
Queue
\end{verbatim}



\newpage
\begin{thebibliography}{99}

\bibitem{hepix-vm-benchmark}
M. Alef and I. Gable.
HEP specific benchmarks of virtual machines on multi-core CPU architectures.
J. Phys Conf. Ser. 219, 052015 (2008).  
doi: 10.1088/1742-6596/219/5/052015

\bibitem{chep-vm}
A. Agarwal, A. Charbonneau, R. Desmarais, R. Enge, I. Gable, D. Grundy, 
D. Penfold-Brown, R. Seuster, R.J. Sobie, and D.C. Vanderster.
Deploying HEP Applications Using Xen and Globus Virtual Workspaces. 
J. Phys.: Conf. Ser. 119, 062002 (2008).
 doi: 10.1088/1742-6596/119/6/062002

\bibitem{xen}
P. Barham, B. Dragovic, K. Fraser, S. Hand, T. Harris, A. Ho, R. Neugebauer, 
I. Pratt  and A. Warfield.
Xen and the art of virtualization.
SOSP03: Proceedings of the Nineteenth ACM Symposium on Operating Systems Principles.

\bibitem{kvm}
KVM (Kernel Based Virtual Machine).
http://www.linux-kvm.org/page/Main\_Page.

\bibitem{nimbus}
K.Keahey, I.Foster, T.Freeman, and X. Zhang.
Virtual workspaces: Achieving quality of service and quality of life in the Grid. 
Sci. Program. Vol. 13 (2005) 265.

\bibitem{one-click-cluster}
K.Keahey and T.Freeman.
Contextualization: Providing One-Click Virtual Clusters.
SCIENCE08: Proceedings of the 2008 Fourth IEEE International Conference on eScience.

\bibitem{open-nebula}
Open Nebula (Open Source Toolkit for Cloud Computing).
http://www.opennebula.org/.

\bibitem{eucalyptus}
D. Nurmi, R. Wolski, C. Grzegorczyk, G. Obertelli, S. Soman, L. Youseff, and D. Zagorodnov. 
The Eucalyptus Open-Source Cloud-Computing System.
CCGRID '09: Proceedings of the 2009 9th IEEE/ACM International 
Symposium on Cluster Computing and the Grid. 
doi:10.1109/CCGRID.2009.93

\bibitem{condor}
D.Thani, T.Tannenbaum and M.Livny.
Distributed computing in practice: the Condor experience.
Concurrency and Computation: Practice and Experience
Vol. 17 (2005) 323.

\bibitem{rfc3820}
S.Tuecke, V.Welch, D.Engert, L.Pearlman,and M.Thompson.
Internet X.509 Public Key Infrastructure (PKI) Proxy Certificate Profile.
http://www.ietf.org/rfc/rfc3820.txt.

\bibitem{gridx1}
A. Agarwal, M. Ahmed, A. Berman, B.L. Caron, A. Charbonneau, D. Deatrich, 
R. Desmarais, A. Dimopoulos, I. Gable, L.S. Groer, R. Haria, R. Impey, L. Klektau, 
C. Lindsay, G. Mateescu, Q. Matthews, A. Norton, W. Podaima, D. Quesnel, 
R. Simmonds, R.J. Sobie, B.St. Arnaud, C. Usher, D.C. Vanderster, 
M. Vetterli, R. Walker, M. Yuen.
GridX1: A Canadian computational grid.
Future Generation Computer Systems {\bf 23} (2007) 680. 
doi:10.1016/j.future.2006.12.006

\bibitem{babargrid}
A. Agarwal, P. Armstrong, R. Desmarais, I. Gable, S. Popov, S. Ramage, 
S. Schaffer, C. Sobie, R.J. Sobie, T. Sulivan, D.C. Vanderster, G. Mateescu, 
W. Podaima, A. Charbonneau, R. Impey, M. Viswanathan, D. Quesnel, 
BaBar MC Production on the Canadian Grid using a Web Services Approach.
J. Phys.: Conf. Ser. 119 (2007) 072002.
doi: 10.1088/1742-6596/119/7/072002

\bibitem{canarie}
CANARIE. Canada's Advanced Research and Innovation Network.
http://www.canarie.ca.

\bibitem{canfar}
CANFAR. Canadian Advanced Network For Astronomical Research. \\
http://astrowww.phys.uvic.ca/~canfar/

\bibitem{cloudscheduler}
The \cs code is available at 
http://cloudscheduler.org.

\end{thebibliography}
\end{document}